\documentclass[10pt]{article}
\usepackage{amsmath,amssymb,amsfonts}
\usepackage{graphics,graphicx}
\usepackage{hyperref}
\usepackage{color}

\newcommand{\1}{\rm I}
\renewcommand{\d}{\partial}
\newcommand{\sign}{\mathop{\rm sign}\nolimits}

\begin{document}

\begin{titlepage}

\vspace{.8cm}
\begin{center}
{\Large{\bf
The universe formation by a space reduction cascade with random
initial parameters}}
\\
[9mm]

\renewcommand{\thefootnote}{\fnsymbol{footnote}}
{\sc S.G.\ Rubin\footnote{$<$\href{mailto:sergeirubin@list.ru}%
{sergeirubin@list.ru}$>$} {\small and} A.S.\ Zinger\footnote{$<$\href{mailto:zinger.alexey@gmail.com}%
{zinger.alexey@gmail.com}$>$}}
\\
[4mm]

{\em\small  National Research Nuclear University "MEPhI", \\ Kashirskoe sh. 31, Moscow, 115409 Russia}
\\

\vspace{2.3cm}

{\sc Abstract}\\
\end{center}

\begin{quote}

In this paper we discuss the creation of our universe using the
idea of extra dimensions. The initial, multidimensional Lagrangian
contains only metric tensor. We have found many sets of the
numerical values of the Lagrangian parameters corresponding to the
observed low-energy physics of our universe. Different initial
parameters can lead to the same values of fundamental constants by
the appropriate choice of a dimensional reduction cascade. This
result diminishes the significance of the search for the 'unique'
initial Lagrangian. We also have obtained a large number of
low-energy vacua, which is known as a 'landscape' in the string
theory.

\end{quote}

\vspace{1.9cm}


\vfill

\center \today{}

\setcounter{footnote}{0}
\end{titlepage}

\section{Introduction}

In this paper we investigate the problems of an emergence of
fundamental constants as well as other parameters at an early
stage of the Universe evolution. It is well known that the range
of admissible parameters must be extremely narrow ({so-called}
fine tuning of parameters) {for such complex structures as our
Universe to exist}, which is difficult to explain. Extensive
literature is devoted to the discussion of this problem
\cite{Fine}. One way of solving it is based on the assumption of
the multiplicity of universes with different properties
\cite{Landscape0}, \cite{Random}, \cite{Landscape1}. Rich
possibilities for justifying this assumption are contained in the
idea of multidimensionality of {the} space. The number of extra
dimensions has long been a subject of debate. For instance, the
Kaluza-Klein model originally contained one extra dimension. Now
infinite-dimensional spaces \cite{Castro} and even
variable-dimensional spaces \cite{Bleyer} are being discussed.

The idea of the multiplicity of universes \cite{Bousso} usually
implies an existence of the initial Lagrangian with specific
parameters and having potential density with numerous local
minima. Each of these minima corresponds to a certain low-energy
effective theory with its own unique set of parameters. Which of
the minima we get to depends on the initial conditions. Therefore,
observed low-energy physics depends not only on the initial
parameters of the Lagrangian, but also on the initial metric tensor of
the created space-time.

This idea is usually developed within the framework of the string
theory \cite{Susskind}, which  {includes, among other
assumptions}, an existence of extra
dimensions. In this paper we develop a purely geometrical approach
\cite{BR06}, postulating only the existence of additional
dimensions. We use the cascade reduction mechanism, introduced in
\cite{Rubin08}, to explain the fine tuning of the parameters in
our universe without involving the string theory assumptions other
than existence of extra dimensions.

The essence of our idea is as follows. Consider a space of
multiple dimensions. Because of quantum fluctuations in some of
its regions, a geometry of the direct product of two subspaces
may arise. Suppose the curvature of one of the subspaces
significantly exceeds that of the other; let us refer to the
former subspace as the \emph{extra} one and to the latter as the
\emph{main} subspace. Quantum fluctuations in some region of a
newly formed main subspace similarly divide it into direct product
of two subspaces. Considering further divisions we arrive at a
"chain" of partitions of the space.  Every such partition we shall
call a \emph{reduction} of the space. Multiple consecutive
reductions we shall call a \emph{cascade of reductions}.

Thus a \emph{cascade of reductions} consists of several steps,
reducing effective dimensionality of the space. It will be shown
that every step of a cascade changes parameters of Lagrangian.
Therefore, by choosing different cascades we can obtain different
final Lagranigians starting from a fixed initial parameters. Each
Lagrangian corresponds to universes with distinct properties. It
is this possibility that is usually associated with the concepts of
the landscape -- numerous low-energy vacua. Our goal is, in a sense, the opposite one: we
try to assess the set of all the parameters of the initial
Lagrangian leading to the observed fundamental constants.

In this paper we show that there are numerous initial Lagrangians
leading to the observable physics. It will be shown that the
low-energy physics depends not only on the initial parameters, but
also on the properties of the compact spaces in particular
cascade. Therefore a variation of initial parameters may be
compensated by an appropriate variation of the properties of the
cascade, leaving the low-energy physics unchanged. This diminishes the importance of the search for the "unique"\ Lagrangian of the Theory of Everything.

We also discuss the relationships between the parameters of the initial Lagrangian and fundamental constants  $\hbar$ and $G$,
determined in low-energy experiments.

\section{Reduction cascades}

\subsection{Main idea} \label{Main}

Let us discuss the main idea
in detail. Consider a $D$-dimensional space $M_D$. Because of
{its} metric fluctuations, new spatial regions with various
geometries are constantly born within it. We'll be interested in
regions with the direct product geometry of the form
\begin{equation}\label{initmetrics}
U_1 =T \otimes M_{d_1} \otimes M_{D_1}, \qquad U_1 \subset M_D
\end{equation}
Here $T$ is the timelike direction,  $M_{d_1}$ is a
compact space of $d_1$ dimensions, $M_{D_1}$ is the main space,
whose metric fluctuations are studied at the next step of the
cascade. We emphasize that geometry is changed not in the whole
space $M_D$, but only in its small region $U_1$. All physical
processes are considered from the viewpoint of an observer located
inside the subspace $M_{D_1}$.

Let us limit our consideration
to {quantum} fluctuations satisfying the following relationship
for Ricci scalars in subspaces $M_{d_1}$ and $M_{D_1}$
\begin{equation}\label{Ricci}
R_{d_1} >> R_{D_1}.
\end{equation}

In case of the simplest geometries, the
bigger the curvature of the space, the less is its volume. This is
the type of geometries we'll be studying. But the less the volume
of the system, the faster are the relaxation processes in it (see
discussion in \cite{Rubin08}). Therefore, due to condition
\eqref{Ricci}, processes in $M_{d_1}$ advance much faster than
those in $M_{D_1}$. In what follows we discuss the conditions to
stabilize the volume of $M_{d_1}$ and its geometry. Theory becomes
effectively $D_1$-dimensional, with $d_1$ compact extra
dimensions. The parameters of the initial $D$-dimensional
Lagrangian are renormalized, with their new values depending on
the properties of the compact extra space $M_{d_1}$.

Described above is the first step of the cascade. The following
steps are similar: due to the metric fluctuations in some volume
of the newly formed space $M_{D_1}$, there arises a new geometry
\begin{equation}\label{initmetrics2}
U_2 =T \otimes M_{d_2} \otimes M_{D_2}, \qquad U_2 \subset
T\otimes M_{D_1}
\end{equation}
resulting in segregation of another compact space $M_{d_2}$. Ricci scalars of both new subspaces should satisfy the condition, analogous to \eqref{Ricci}:
\begin{equation}\label{Ricci2}
R_{d_2} >> R_{D_2},
\end{equation}
Parameters of the reduced Lagrangian are renormalized once more; their values become dependent on the properties of the compact extra space $M_{d_2}$.

Cascades differ from each other by the properties of its
compact subspaces: their volume, topology and geometry.

Each reduction consists of two stages: first is the quantum formation of the space of the form \eqref{initmetrics}; second is the classical evolution of this space, which results in stabilization of the compact extra space $M_{d_i}$.

\subsection{The first step of the cascade}

\subsubsection{Quantum formation of the space}\label{QC}
In this section we discuss the probability of a quantum formation
of geometries we are interested in. Let us consider a
$D$-dimensional space. Quantum fluctuations in small regions of
this subspace create subspaces of the form $M_{d_1} \otimes
M_{D_1}$. 

{The transitions with a change in the geometry are conveniently described in terms of the path integral technique}
\cite{HH,V2}. For this purpose, the superspace $\mathcal M_D =(M_D ; g_{ij})$ is defined as a set of metrics $g_{ij}$ in the space $M_D$ to within diffeomorphisms. On a spacelike section
$\Sigma$ we introduce the metric $h_{ij}$ (for details, see \cite{Wilt}) and
define the space of all Riemannian $(D-1)$ metrics in
the form
\begin{equation*}
Riem(\Sigma ) = \{ h_{ij} (x)\left| {x \in \Sigma } \right.\}
\end{equation*}
The transition amplitude from the section $\Sigma _{in}$ to the section $\Sigma _{f}$ is the integral over all geometries allowable by the boundary conditions:
\begin{equation}
A_{f,in}=\left\langle {h_{f},\Sigma _{f}|{h_{in},\Sigma
}_{in}}\right\rangle =\int_{h_{in}}^{h_{f}}Dg\exp [iS(g)] .
\label{Amplitude}
\end{equation}

The absence of the Planck constant in the exponential
function is a result of choosing the appropriate units of
measurement. It will be shown below
that the Planck constant $\hbar$ naturally emerges after the
definition of the dimensional units simultaneously
with gravitational constant $G$. However, the statement
on the unification of gravity and quantum theory
would be premature. The essence of quantum
mechanics is based on the rule of summation of the
transition amplitudes  (\ref{Amplitude}), which is postulated originally.

What is the probability of such a process? The answer is
far from clarity even for the birth of a 4-dimensional space in
the standard theory of gravitation linear in scalar curvature \cite{HH,V2}. As
we need only to verify that this probability is nonzero, let us
approximate the Largrangian by a linear theory. This is ensured by
the condition \eqref{Ricci}.

Action in the space $M_D$ is chosen in the form
\begin{eqnarray}\label{act0}
S_D = N_0 \int {d^D X\sqrt {\left|G^{(D)}\right|} F(R;a_n)};\quad
F(R;a_n) = \sum\limits_n {a_n R^n },\quad a_1 =1,
\end{eqnarray}
where $G^{(D)} \equiv {\rm det} (G_{AB})$, $R$ is a Ricci scalar,
$N_0$ and $\{a_i\}$ are constants (see, for
example, \cite{Star,Nojiri,Sokolowski}). Note that normalization
constant $N_0$ equals $1/(16\pi G)$ only in a low energy limit. Let us refer to $S_D$ as the first generation action. The standard form is $F(R;a_n)=R-2\Lambda$, i.e. $a_1=1,\quad a_0 = -2\Lambda$.

First step of a cascade begins in a small volume region with forming a subspace
\eqref{initmetrics} with the metric of the form
\begin{eqnarray}\label{interval}
ds^{2}&=&G_{AB}dX^{A}dX^{B}={\rm{N}}dt^2  - g_{ab}(x) dx^{a}dx^{b}-
e^{2\beta(x)}\gamma_{ij}(y)dy^{i}dy^{j}.
\end{eqnarray}
where  $g_{ab}(x)$ is a spatial part of metric in $M_{D_1}$,
$\gamma_{ij}(y)$ is a positively defined metric of the extra space
$M_{d_1}$ and $e^{2\beta(x)}$ is a scaling factor (see
\cite{Carroll,Brokoru07}). Metric \eqref{interval} is the generalisation of minisuperspace geometry, see e.g. \cite{V1}
\begin{eqnarray}\label{minisuper}
&&ds^{2}=\sigma ^{2}\left[ N(t)^{2}dt^{2}-a(t)^{2}d\Omega
_{D_1}^{2}\right].
\end{eqnarray}
Here $N(t)$ is a lapse function, and $a(t)$ is a scale
factor. The factor $\sigma ^{2}$ is usually expressed as $\sigma
^{2}=1/(12\pi ^{2}M^{2}_{Pl})$ when dealing with 4-dimensional
gravity.

Let us find the probability of producing such a space, approximating it by a linear dependence on $R$. Note that owing to the
form of the chosen metric \eqref{interval}, the following
relationships hold true (see \cite{BR06}):
\begin{eqnarray}                                   \label{R-decomp}
      &&  R_D = R_{D_1} + R_{d_1}(\gamma_{ij}) + f_{\rm der},
\nonumber \\
      &&  \phi \equiv R_{d_1}(\gamma_{ij})= kd_1 (d_1 -1) e^{-2\beta(x)}
 \\
      && f_{\rm der} = 2d_1 g^{ab}\nabla_{a}\nabla_{b}\beta
                + d_1 (d_1 +1) g^{ab}\partial_{a}\beta \partial_{b}\beta, \nonumber
\end{eqnarray}
The covariant derivative $\nabla$ acts in the space $M_{D_1}$.

As was shown in \cite{Rubin09}, an extra space with an arbitrary
geometry evolves into a space with a maximum number of Killing
vectors for a given topology. So we can choose a maximally
symmetric space with a constant Ricci scalar $R_{d_1}$ and a
curvature $k=\pm 1$.

The volume $V_{d_1}$ of an internal space of a
unit curvature depends on its geometry, because it is given by its intrinsic metric
\begin{equation} \label{volume}
    V_{d_1}=\int d^{d_1}y\sqrt{\left|G^{(d_1)}\right|}.
\end{equation}

Let us emphasize that it would suffice to demand that only a small fracture of the volume of the initial space $M_D$ has the geometry \eqref{interval}. The following discussion is concerned with the point of view of an internal observer within $M_{D_1}$, who cannot get any information from the outside.

We will use the slow-change approximation
proposed in \cite{BR06}:
\begin{equation}\label{ineq}
|R_{d_1}| \gg |R_{D_1}|,\quad |R_{d_1}|\gg|f_{der}|
\end{equation}

Substituting Eq.\eqref{R-decomp}
into Eq.\eqref{act0} yields
\begin{equation}\label{FTailor}
F(R_D) = F(R_{D_1} + R_{d_1} + f_{\rm der})\simeq F(R_{d_1}) +
F'(R_{d_1})R_{D_1} + F'(R_{d_1})f_{\rm der}
\end{equation}
Also, determinants satisfy
\begin{equation}\label{det2}
\left|G^{(D)}\right| =e^{2\beta(x)} \cdot \left|G^{(D_1)}\right| \cdot \left|G^{(d_1)}\right|.
\end{equation}
Substituting these expressions into action \eqref{act0} and
carrying out certain computations given explicitly in \cite{BR06},
we obtain a Lagrangian of a scalar-tensor gravitation in $D_1$
dimensions
\begin{eqnarray}\label{E-H_ActionD1}
     &&S_{{D_1}} = N'_0 \int d^{D_1}x\, \sqrt{ \left|G^{(D_1)}\right|}\, (\sign F') [R_{D_1} + \frac{1}{2} K(\phi) (\partial\phi)^2 - U(\phi)],
%
\\
     &&K(\phi) =                                \label{KE}
        \frac{1}{2\phi^2} \left[
            6\phi^2 \biggl(\frac{F(\phi;a_n)''}{F(\phi;a_n)'}\biggr)^2\!
            -2 d_1 \phi \frac{F(\phi;a_n)''}{F(\phi;a_n)'} + \frac{1}{2} d_1 (d_1 + 2)\right],
\\
     &&U(\phi) = - (\sign F(\phi;a_n)')
        \left[\frac{|\phi|}{d_1 (d_1 -1)}\right]^{d_1/2}
                \frac{F(\phi;a_n)}{F'(\phi;a_n)^2 }    .            \label{VE}
\end{eqnarray}
In accordance with \eqref{ineq}, we have kept only the term linear in Ricci scalar.

The quantum birth of the Universe in the linear theory has been studied by many authors. It is usually examined within the framework of minisuperspace, where the interval is written in the form \eqref{minisuper}

 The {probability of} the $D_1$-{dimensional} space quantum birth was calculated in
\cite{Carlip}. The universe creation probability in the presence of a scalar field has been studied in many papers, {for instance \cite{V2}}. In Vilenkin's
approach, the probability of a 3-dimensional space birth is
$dP\propto \exp\left[{\frac{+2}{3U(\phi)}}\right], $ while the
Hartle-Hawking approximation yields $dP\propto
\exp\left[{\frac{-2}{3U(\phi)}}\right] $.
For all their differences, the main result of both approaches is a
nonzero probability of such an event. Hence, the fraction of the
universes with given properties produced by a cascade of
reductions is nonzero.

\subsubsection{Stage of the classical evolution}

In order to estimate the probability of the quantum birth of a space consisting of two subspaces -- the main one and the additional one -- we restricted the discussion in previous section \ref{QC} to a linear in scalar curvature approximation.

{Let us confine ourselves to those initial parameters,  which {lead to potentials having minima}. Our numerical calculations indicate that such parameters do exist.}

After nucleation, classical dynamics of these subspaces is as follows.
A compact extra subspace evolves in such a way that the field
$\phi$ approaches the value $\phi_m$, corresponding to the minimum
of the potential $U(\phi)$.

It follows from the
definition of the field $\phi$ \eqref{R-decomp} and the expression
for the interval \eqref{interval} that the characteristic size of
the space $M_{d_1}$ is proportional to $e^{\beta}\propto
1/\sqrt{\phi}$. Therefore, the size of a compact \emph{extra}
space quickly stabilizes when the field reaches the value $\phi =
\phi_m$. This corresponds to $R_{d_1}\rightarrow const,\quad
f_{der}\rightarrow 0$. Therefore, in the following discussion we
can use the conditions
\begin{equation}\label{stat}
R_{d_1}=const,\quad f_{der}=0,
\end{equation}
which significantly simplify the calculations.

The equation for the scaling factor of the main subspace during de
Sitter stage is of the form \cite{Carroll}
\begin{equation}\label{scalef}
\frac{D_1 (D_1 -1)}{2}(\dot{a}/a)^2 = \Lambda - \frac{D_1 (D_1
-1)}{2}k
\end{equation}
We assume that the size of an extra space $M_{d_1}$ has stabilized and $U(\phi) \simeq U(\phi_m )\equiv\Lambda$. Consequently, the scaling factor depends on time as
$$a(t)\propto e^{Ht}, \quad H=\frac{2\Lambda}{D_1 (D_1 -1)}$$
for large $t$. The size of a main subspace rapidly increases.

Thus, we have $D_1$-dimensional quickly expanding space and $d_1$-dimensional compact extra space. Linear in curvature approximation was sufficient to obtain this result, but to advance further we will need a more accurate expression for the reduced action. The latter could be derived using conditions \eqref{stat}. Expanding formula $F(R; a_n) = F(R_{d_1} + R_{D_1};a_n)$
into Taylor series and integrating over coordinates of extra
dimensions, we obtain
\begin{eqnarray}\label{act2}
S_{D_1} = N'_0 \int {d^{D_1} X\sqrt{\left|G^{(D_1)}\right|}F(R_{D_1};\tilde{a}_n)}.
\end{eqnarray}
Here new parameters $\tilde{a}_n$ are functions of $R_{d_1}$, $n>0$.

We arrive at the second generation of action \eqref{act2}, which is similar to the first generation of action \eqref{act0} with changed numerical values of the parameters $\tilde{a}_n$. Dimensionality of the main space has been reduced, $D_1 <D$, and a new compact extra space of $d_1 =D-D_1$ dimensions has been formed.

The size of the space $M_{D_1}$ is significantly larger than the size of $M_{d_1}$. In this paper we are concerned only with quantum fluctuations creating spaces satisfying such relation. Subsequent dynamics further increases their disparity.

The second step of a cascade is analogous to the first one with substitution $S_D \to S_{D_1}$: in a small region of the space $M_{D_1}$ there occurs a quantum fluctuation which creates a subspace with topology $M_{D_2}\otimes M_{d_2};\quad D_2+d_2 =D_1$. As a result of classical dynamics, the size of the space $M_{d_2}$ is stabilized and the space  $M_{D_2}$ is expanding.

If we wish not to be concerned with excitations of the compact space $M_{d_1}$, we should only consider such quantum fluctuations in $M_{D_1}$ that satisfy $R_{d_2}<<R_{d_1}$.

\section{Quadratic gravity as an explicit example} \label{Quadratic}

Initial action for gravitational field includes all powers of Ricci scalar and other invariants. A vast majority of the works on the subject uses some finite polynomial in Ricci scalar. A choice of a particular polynomial may be justified as follows. Let us consider a quantum fluctuation that produces a geometry with characteristic value of a scalar curvature $R_0$. Then the initial Lagrangian \eqref{act0} may be approximated by a finite polynomial:
\begin{equation}\label{approx}
F(R,a_n)\simeq \sum\limits_{k=-K}^{K'} {b(R_0 )_k (R-R_0)^k }
\end{equation}
Specific values of $K$ and $K'$ are designated according to the author's purposes.
Coefficients $b_k(R_0 )$ depend on the location of
the expansion \eqref{approx} and vary over a wide range.

Consider quadratic gravity in the space $M_D$ with an action of
the form
\begin{align} \label{S1}
     S_{D}=\frac{N_0}{2}&\int d^{D}X\sqrt{\left|G^{(D)}\right|}
        \bigg[  R_{D}(G_{AB})+CR_{D}^2(G_{AB}) +C_1 R_{AB} R^{AB}+  \nonumber \\
&+C_2 \mathcal{K} - 2 \Lambda\bigg]+ \int\limits_{\partial M_D} Kd^{D-1}\Sigma,
\end{align}
where we also included Ricci tensor and Kretschmann scalar
$\mathcal{K}=R_{ABCD} R^{ABCD}$ {The
boundary term ${(\partial M_D)}$ introduced by Hawking and Gibbons
does not influence classical dynamics and we ignore it in the
following \cite{Carlip}.}

Let us try to find the values of the parameters that allow
a universe similar to ours to form. In this case a set of
parameters of the Lagrangian $\{a_n\}$ (see \eqref{act0}) is
$\{C,C_1,C_2,\Lambda \}$.

Following the steps outlined in section  \ref{Main}, we will find
the form of an action \eqref{S1}, reduced to space $M_{D_1}$.
{An action pertaining only to $M_{D_1}$ can be recovered by
integrating action in $M_D$ \eqref{S1} over $M_{d_1}$}.

{Ricci scalar can be expressed using equations}
\eqref{R-decomp}:
\begin{equation}
R_D(G_{AB})=R_{D_1}(g_{ab})+\phi(x) \label{R-decomp2}
\end{equation}
where we have set $f_{der}=0$ - see \eqref{stat}.  The
decomposition of $R_{AB}R^{AB}$ and $\mathcal{K}$ is given by (see
\cite{BR06})
\begin{align}
R_{AB}R^{AB}&=R_{ab}R^{ab}+e^{-4\beta(x)}\cdot R_{\mu\nu}R^{\mu\nu} \nonumber \\
\mathcal{K}&=\mathcal{K}(g_{ab})+e^{-4\beta(x)}\cdot\mathcal{K}(\gamma_{ij}), \label{K-decomp}
\end{align}
where variables with indexes $A,B$ correspond to metric $G_{AB}$,
those with indexes $a,b$ correspond to $g_{ab}$, and indexes $\mu,
\nu$ correspond to metric $\gamma_{ij}$, (i.e. those values
correspond to spaces $M_{D}$, $M_{D_1}$ and $M_{d_1}$
respectively).

To advance further, recall that we are considering the
$d_1$-dimensional space metric $\gamma_{ij}$ of a constant
curvature $k$, so that we can express the Riemann tensor, Ricci
tensor and Ricci scalar in ${M}_{d_1}$ space in terms of its
curvature:
\begin{align} \label{r0a}
    R^{\mu\nu}{}_{\rho\eta} = k\, \delta^{\mu\nu}{}_{\rho\eta};
\qquad\quad
    R_\mu^\nu = k\,  (d_1-1) \delta_\mu^\nu;
\qquad\quad
    R_{d_1}' \equiv k\,  d_1 (d_1-1)
\end{align}
where $\delta^{\mu\nu}{}_{\rho\eta} = \delta^\mu_\rho
\delta^\nu_\eta-\delta^\mu_\eta \delta^\nu_\rho$. $R_{d_1}'$
represents the characteristic scale of curvature of extra
dimensions. Expressions for squared Riemann and Ricci tensors are
derived from \eqref{r0a}; the former is by definition a
Kretschmann scalar:
\begin{align}
R_{\mu\nu}R^{\mu\nu}&=d_1[k(d_1-1)]^2
\nonumber \\
\mathcal{K}(\gamma_{ij})\equiv R_{\mu\nu\rho\eta}&R^{\mu\nu\rho\eta}=2d_1(d_1-1)k^2 \label{R,K,2}
\end{align}

Now we can rewrite \eqref{K-decomp} substituting $e^{\beta(x)}$
from \eqref{R-decomp} and using \eqref{R,K,2}:
\begin{align}
R_{AB}R^{AB}&=R_{ab}R^{ab}+\frac{1}{d_1}\phi(x) \nonumber \\
\mathcal{K}=\mathcal{K}&(g_{ab})+\frac{2}{d_1(d_1-1)}\phi(x), \label{K-decomp2}
\end{align}
\newline
After plugging \eqref{R-decomp2} and \eqref{K-decomp2} into action
\eqref{S1} and grouping the terms we obtain
\begin{align}
S_D=\frac{N_0}{2} \int
d^Dx\sqrt{\left|G^{(D)}\right|}\times\Big\{&R_{D_1}(g_{ab})(1+2C\phi)+CR_{D_1}^2(g_{ab})-2\Lambda+
\nonumber \\
+C_1R_{ab}R^{ab}+ C_2\mathcal{K}(g_{ab}&)+\phi +
\left(C+\frac{C_1}{d_1}+\frac{2C_2}{d_1(d_1-1)}\right)\phi^2\Big\} \label{actF} =
\nonumber \\
\qquad = \frac{N_0}{2} \int d^Dx\sqrt{\left|G^{(D)}\right|} \cdot\mathcal{L}&(g_{ab})
\end{align}
The value in brackets is denoted $\mathcal{L}(g_{ab})$ for
convenience; it does not depend on the coordinates of
extra space $M_{d_1}$.

To find the action in space $M_{D_1}$ we'll have to integrate
\eqref{actF} over the space $M_{d_1}$ using definition of the
volume of that space \eqref{volume}. Substituting relationship
\eqref{det2} for $\sqrt{\left|G^{(D)}\right|}$ and relationships
\eqref{R-decomp} and \eqref{r0a} for $e^{2\beta(x)}$ yields:
\begin{align}
S_{D}&=\frac{N_0}{2}\int d^Dx \sqrt{\left|G^{(D)}\right|} \cdot\mathcal{L}(g_{ab})= \nonumber \\
&= \frac{N_0}{2}\int d^{D_1}x \ e^{2\beta(x) d_1}\sqrt{\left|G^{(D_1)}\right|} \cdot\mathcal{L}(g_{ab}) \times \int d^{d_1}x \sqrt{\left|G^{(d_1)}\right|} = \nonumber \\
&= \frac{N_0 V_{d_1}}{2} \int d^{D_1} x \sqrt{\left|G^{(D_1)}\right|} \left(\frac{R_{d_1}'(G^{(d_1)})}{\phi(x)}\right)^{d_1/2}   \cdot\mathcal{L}(g_{ab})       \label{S2}
\end{align}

Let us suppose that the minimum of a potential $U(\phi)$
exists. The field $\phi(x)$ rapidly relaxes to it and stays fixed
during the low-energy processes (see \cite{Rubin09} for
discussion). This case is the most natural, since the relaxation
time is proportional to the scale of the extra space $M_{d_{1}}$,
which is small compared to the scale of the space $M_{D_{1}}$.

Assuming those conditions are satisfied, let us perform a
conformal transformation of the form (see, for instance,
\cite{Bron})
\begin{align}
&g_{ab} = | f(\phi_{m})|^{-2/(D_1-2)} \,\widetilde{g}_{ab}, \qquad \qquad f(\phi)\equiv \phi^{-d_1/2}(x)\left[1+2C\phi(x)\right],  \label{f}
\\
   & R_{D_1}  = | f(\phi_{m})|^{2/(D_1-2)}\widetilde{R}_{D_1},
\nonumber \\
&\mathcal{K}=| f(\phi_{m})|^{8/(D_1-2)}\widetilde{\mathcal{K}}
\nonumber \\
       &\sqrt{\left|G^{(D_1)}\right|} = | f(\phi_{m})|^{-D_1/(D_1-2)} \sqrt{|\widetilde{G}^{(D_1)}|}, \nonumber
\end{align}
which, being applied to Eq. \eqref{S2}, brings us to the initial
form of the action (compare with \eqref{S1}):
\begin{align}
    S_{D_1}  =& \frac{N_0^1}{2}\int d^{D_1}x\sqrt{\left|G^{(D_1)}\right|}\Big\{R_{D_1}(g_{ab}) +C^{(D_1)}R_{D_1}(g_{ab})^2 +
\nonumber \\
&+C^{(D_1)}_1R_{ab}R^{ab}(g_{ab})+C^{(D_1)}_2\mathcal{K}(g_{ab}) -2\Lambda^{(D_1)}\Big\}, \label{S4}
\end{align}
where the tildes were omitted for short.
New parameters are expressed in terms of the old ones by the
following equations:
\begin{align}
    C^{(D_1)}
    = \sign& \left(f(\phi_m)\right) \ |f(\phi_{m})|^{(4-D_1)/(D_1-2)}
    \phi_{m}^{-d_1/2}C,             \label{c4}
 \\
    \Lambda^{(D_1)}
    = \sign &\left(f(\phi_m)\right)  |f(\phi_{m})|^{-D_1/(D_1{-}2)}
        \phi_{m}^{-d_1/2} \times \nonumber \\
    &\times \left[\Lambda-1/2 \left(\phi_{m}{+}\left(C+\frac{C_1}{d_1}+\frac{2C_2}{d_1(d_1-1)}\right)\phi_m^2 \right) \right], \\
C^{(D_1)}_1
    = \sign& \left(f(\phi_m)\right) \  |f(\phi_{m})|^{(4-D_1)/(D_1-2)}
    \phi_{m}^{-d_1/2}C_{1},
\\
C^{(D_1)}_2
    = \sign& \left(f(\phi_m)\right) \ |f(\phi_{m})|^{(8-D_1)/(D_1-2)}
    \phi_{m}^{-d_1/2}C_{2}. \label{c4_2}
\end{align}
where $f(\phi)$ is defined by \eqref{f}. Recall that we are
considering the case where the field $\phi$ is already at its
minimum, $\phi =\phi_m$, so the kinetic terms were neglected.
Equations \eqref{c4}-\eqref{c4_2} connect old and new parameters after one reduction.

Next reduction leads to analogous formulas for parameters
$C^{(D_2)} , C^{(D_2)}_1$, $C^{(D_2)}_2 , \Lambda^{(D_2)}$ with
substitutions $C\rightarrow C^{(D_1)} , C_1\rightarrow C^{(D_1)}_1
, C_2 \rightarrow C^{(D_1)}_2 , \Lambda \rightarrow
\Lambda^{(D_1)}$. Thus, we have obtained recurrence formulas for the
parameters.

Action \eqref{S4} for a subspace $M_{D_1}$ coincides in the form
with the initial one \eqref{S1} for $M_D$, but with renormalized
parameters
$C^{(D_{1})},C^{(D_1)}_1,C_2^{(D_{1})},\Lambda^{(D_{1})}$.

\subsection{The final reduction. Formation of the low energy physics.}
It was shown in the previous section that the form of the action
\eqref{S1} does not change after space reductions
\eqref{interval}. Now we shall consider the final reduction of an arbitrary cascade of $(n+1)$ reductions -- the reduction leading to a four-dimensional space-time.

It was shown in \cite{BR06} that the action \eqref{S1} may be
reduced to the four-dimensional form
\begin{align}
    S&\simeq \frac{N_0 V_d}{2}\int d^{4}x
        \sqrt{\left|G^{{(4)}}\right|}[R_4 + K(\phi)(\partial \phi)^2
        -2U(\phi)] \label{E-H_Action}
\\
U&(\phi) = -\frac{1}{2}\sign(1 + 2C^{(D_n)} \phi) \cdot \left[\frac{|\phi |}{d_1 (d_1 -1)}\right]^{d_1/2} \frac{C^{(D_n)}_{tot} \phi^{2}+\phi -2\Lambda}{(1+2C^{(D_n)}\phi )^2} \label{U_Ein}
\\
K&(\phi)= \frac{ (C^{(D_n)})^2 \phi^2 (d_1^2 - 2d_1 + 12)  + C^{(D_n)} d_1^2 \phi + \frac{1}{4} d_1 (d_1 +2)}{(1 + 2C^{(D_n)}\phi)^2 \phi^2 }+ \nonumber \\ & \qquad \qquad + \frac {C^{(D_n)}_1 + C^{(D_n)}_2} {2\phi (1 + 2C^{(D_n)}\phi)} \label{K_Ein}
\\
\text{where}& \; C^{(D_n)}_{tot}\equiv C^{(D_n)}+\frac{C^{(D_n)}_1}{d_1}+\frac{2C^{(D_n)}_2}{d_1(d_1-1)} \nonumber
\end{align}
(compare with \eqref{KE},\,\eqref{VE}, which are written for the
case $C_1^{(D)}=C_2^{(D)}=0$). These
equations, though derived for the action \eqref{S1}, are still
applicable after any number of reductions, because the
reductions do not change the form of the action, as was shown in the previous section.

In the vicinity of the minimum, $U(\phi_m)\equiv
\min\big(U(\phi )\big)$, potential can be expanded in a Taylor series so
that the action \eqref{E-H_Action} becomes
\begin{align} \label{S`}
S \simeq \frac{{N_0 V_d }}{2}\!\int&\! d^4 x
{\sqrt{\left|G^{(4)}\right|}}\bigg[ {R_4 + K (\phi_m) (\d \phi )^2
- 2U (\phi _m ) -  U'' (\phi _m ) (\phi {-} \phi_m)^2 }-
\nonumber \\
&-\frac{1}{3}U'''(\phi_m)(\phi-\phi_m)^3 -\dots\bigg].
\end{align}

The unit of measurement \1 is still arbitrary. Let the units be
related by expressions
\begin{equation}\label{Units}
{\1} =\alpha \cdot cm,
\end{equation}
where $\alpha$ is as yet an unknown parameter. We change over to the standard units of length:
\begin{align}\label{ActionIcm}
S = N_0 V_d c \alpha^2 \int & dt d^3 x \bigg[R_4 + {\frac{1}
{2}K(\phi _m )(\partial \phi)^2 - \alpha^2 U(\phi_m ) - \alpha^2
\frac{1}{2}U'' (\phi _m )(\phi - \phi _m )^2 } -
\nonumber \\
&-\alpha^2 \frac{1}{6}U'''(\phi_m)(\phi-\phi_m)^3 \bigg]  .
\end{align}
Here, $x \rightarrow\alpha x$, ${\sqrt{\left|G^{(4)}\right|}}=c_I
=\alpha c$, $R_4 \rightarrow R_4 /\alpha^2$ and $(\partial \phi)^2
\rightarrow (\partial \phi)^2 / \alpha^2$.

{Since expression \eqref{ActionIcm} is the low-energy limit of
action \eqref{S1}, it should adequately describe purely
gravitational phenomena. {Moreover, expression
\eqref{ActionIcm} makes it possible to explain the origin of the
inflaton potential and the cosmological constant.}} The effective
action \eqref{ActionIcm} contains the initial parameters of the
theory without involving fundamental constants, such as the Planck
constant $\hbar$ and the gravitational constant $G$.
{Now we will find the relations between those values.}

Let us introduce the definition
\begin{equation}\label{mass}
m_{\Phi}\equiv \alpha \sqrt{\frac{ U'' (\phi _m )}{K (\phi _m )}}
\end{equation}
and, under the assumption $K(\phi _m )>0$, introduce the
variable
$$\Phi=\sqrt{\frac{c^4}{16\pi G} K(\phi _m )}(\phi  - \phi _m ) .$$
The action \eqref{ActionIcm} in the terms of $\Phi$ is
\begin{align} \label{S_th}
S=\frac{16\pi G}{c^4} N_0 V_d c\alpha^2 \int & dt d^3x \Bigg[ {\frac{c^4}{16\pi G}R_4  + \frac{1}
{2}(\partial \Phi )^2  - \frac{c^4}{16\pi G}\alpha^2 U(\phi_m)  - \frac{1}{2}m_\Phi ^2 \Phi ^2 }
 \nonumber \\
&-\frac{1}{6}\sqrt{\frac{16\pi G}{c^4}}m_\Phi^2 \frac{U'''(\phi_m)}{U''(\phi_m)\sqrt{K(\phi_m)}}\Phi^3 \Bigg]
\end{align}
We arrive at the conventional form of the action for the scalar
field $\Phi$ interacting with gravity:
\begin{equation}\label{actinOrdin}
S = \frac{1}{\hbar}\int dtd^3 x \left( {\frac{c^4}{16\pi G}R_4  +
\frac{1} {2}(\partial \Phi )^2  - \Lambda  - \frac{1}{2}m_\Phi ^2
\Phi ^2 } + \lambda_3\Phi^3 \right) .
\end{equation}
where $G$ is the gravitational constant, $c$ is the speed of
light, $\hbar$ is the Planck's constant, $\Lambda$ is the dark
energy density and $m_\Phi$ is the mass of the scalar field which
we associate with inflaton. 

Comparing \eqref{actinOrdin} with \eqref{S_th} we obtain the
relations
\begin{align}
&\frac{1}{\hbar}=\frac{16\pi G}{c^3} N_0 V_d \alpha^2 \label{hbar} \\
&\Lambda=\frac{c^4}{16\pi G}\alpha^2 U(\phi_m) \label{Lambda2} \\
&\lambda_3 =-\frac{1}{6}\sqrt{\frac{16\pi G}{c^4}}m_\Phi^2
\frac{U'''(\phi_m)}{U''(\phi_m)\sqrt{K(\phi_m)}} \label{la3}
\end{align}

By eliminating the parameter $\alpha$ from relationships \eqref{hbar},
\eqref{Lambda2}, \eqref{la3} and \eqref{mass}, we obtain
\begin{align}
\frac{U(\phi _m )K(\phi _m )} {U''(\phi _m )} &= \frac{16\pi G}
{c^4 }\frac{\Lambda }
{m_\Phi ^2 } , \label{connect1} \\
N_0 V_d\frac{K(\phi_m)}{U''(\phi_m)}
&=\frac{c^3}{16\pi  G \hbar m_\Phi^2} , \label{connect2} \\
\frac{U'''(\phi_m)}{U''(\phi_m)\sqrt{K(\phi_m)}}&=-6\sqrt{\frac{c^4}{16\pi
G}}\frac{\lambda_3}{m_\Phi^2} . \label{connect3}
\end{align}

Right sides of equations \eqref{connect1}-\eqref{connect2} contain
fundamental constants while the left ones depend on the initial
parameters. If the field $\Phi$ is associated with the inflaton,
its mass is about $10^{13}$GeV. Parameter $\lambda_3$ defined in
\eqref{la3}, is not yet measured and may be considered to be the
prediction of our approach. {During numerical calculations we kept in mind that $\lambda_3 < 10^{-12} M_{Pl}$. 
It provides  inflation,  which  does not contradict the observational data.}

The conversion factor between the units of length $\alpha =\1 /cm$
can also be found by another method. Indeed, the characteristic
size of the extra space is determined to be ${V_d}^{1/d}$ in units
of $\1$. Moreover, by designating the size of the hypothetical
extra space as $L_{d}$ expressed in terms of cm, we obtain $\alpha
= L_{d}/{V_d}^{1/d}$. Values of $L_{d} \leq 10^{-17}$ do not
contradict experimental data. Taking into account expression \eqref{mass}, we find the
constraint on the initial parameters
\begin{equation}
L_d = m_\Phi V_d^{1/d} \sqrt{\frac{{K(\phi _m )}} {{U''(\phi_m )}}} < 10^{ - 17}. \label{limit}
\end{equation}

Relationships \eqref{connect1}-\eqref{limit} allow us to compare
the initial parameters of the theory - {$N_0,C,C_1,C_2,\Lambda$}
from the action \eqref{S1}  with observed constants
$c,\hbar,G,\Lambda,m_{\Phi}$ and to predict the value of inflaton
coupling constant $\lambda_3$. The question is whether we can find
the initial parameters to satisfy the constraints
\eqref{connect1}-\eqref{limit}. We have proved the existence of
such parameters by numerical simulations of reduction cascades.
{As it is shown in the next section, we }have found numerous sets
of initial parameters satisfying those conditions.

\subsection{Numerical computations}
The numerical simulations are carried out in the following manner:
first, we assign random values to initial parameters of cascade --
$C,C_1,C_2,\Lambda$. Then we choose a cascade by specifying a
sequence of dimensions of hyperspaces undergoing reductions, which
should end with a four-dimensional space-time. For example,
$\langle 15 \to 11 \to 8 \to 4 \rangle$ constitutes a cascade
going from a 15-dimensional space to a 4-dimensional one. Then we
calculate the values of the final parameters using recurrence
formulas \eqref{c4}-\eqref{c4_2}; substituting them into
\eqref{U_Ein},\eqref{K_Ein} we obtain kinetic and potential terms
before the last reduction. Those final parameters which satisfy
the conditions \eqref{connect1}-\eqref{limit}, correspond to
observable low-energy physics.

\begin{figure}
\centering
\includegraphics[width=0.49\textwidth]{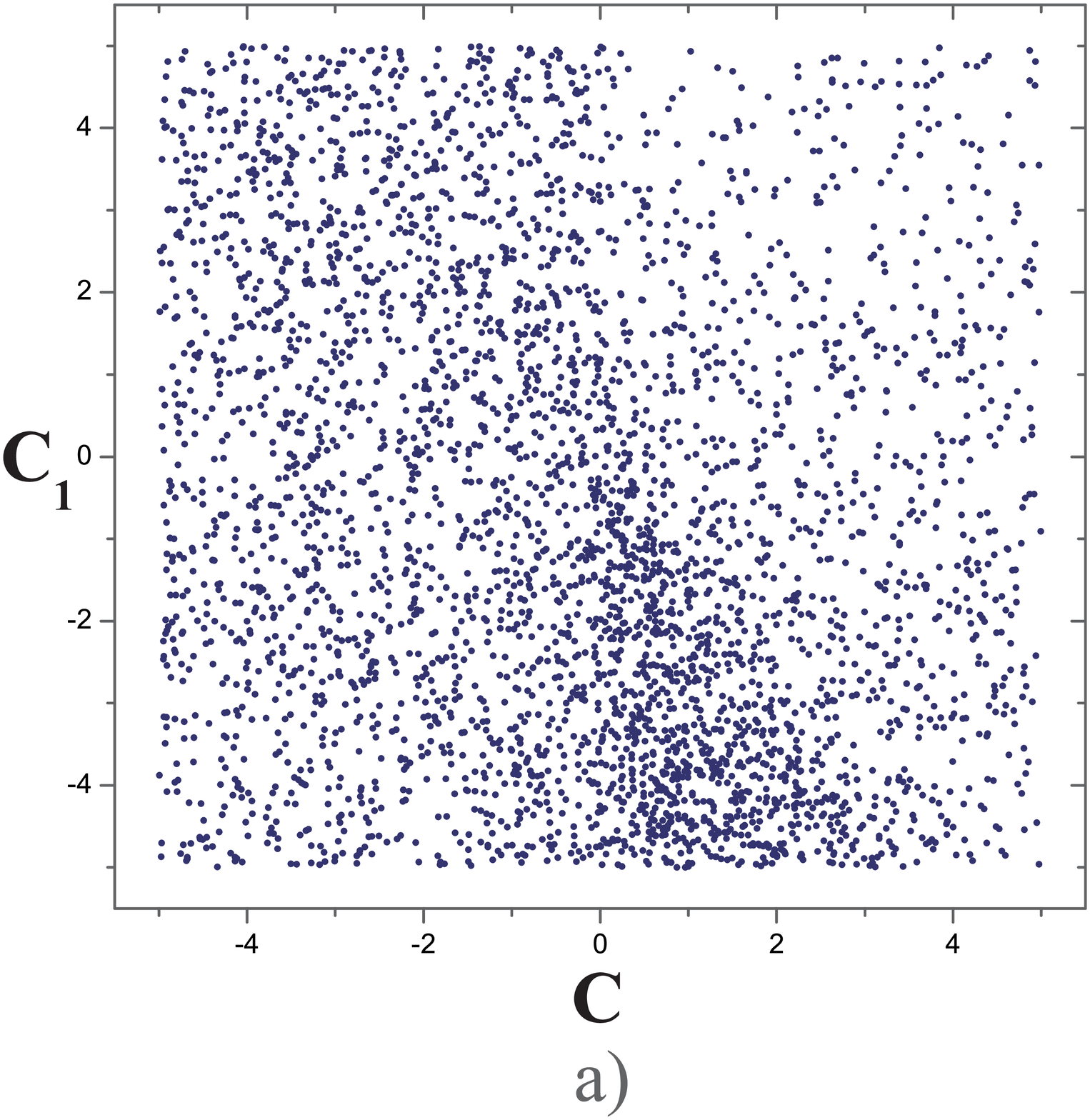}
\includegraphics[width=0.49\textwidth]{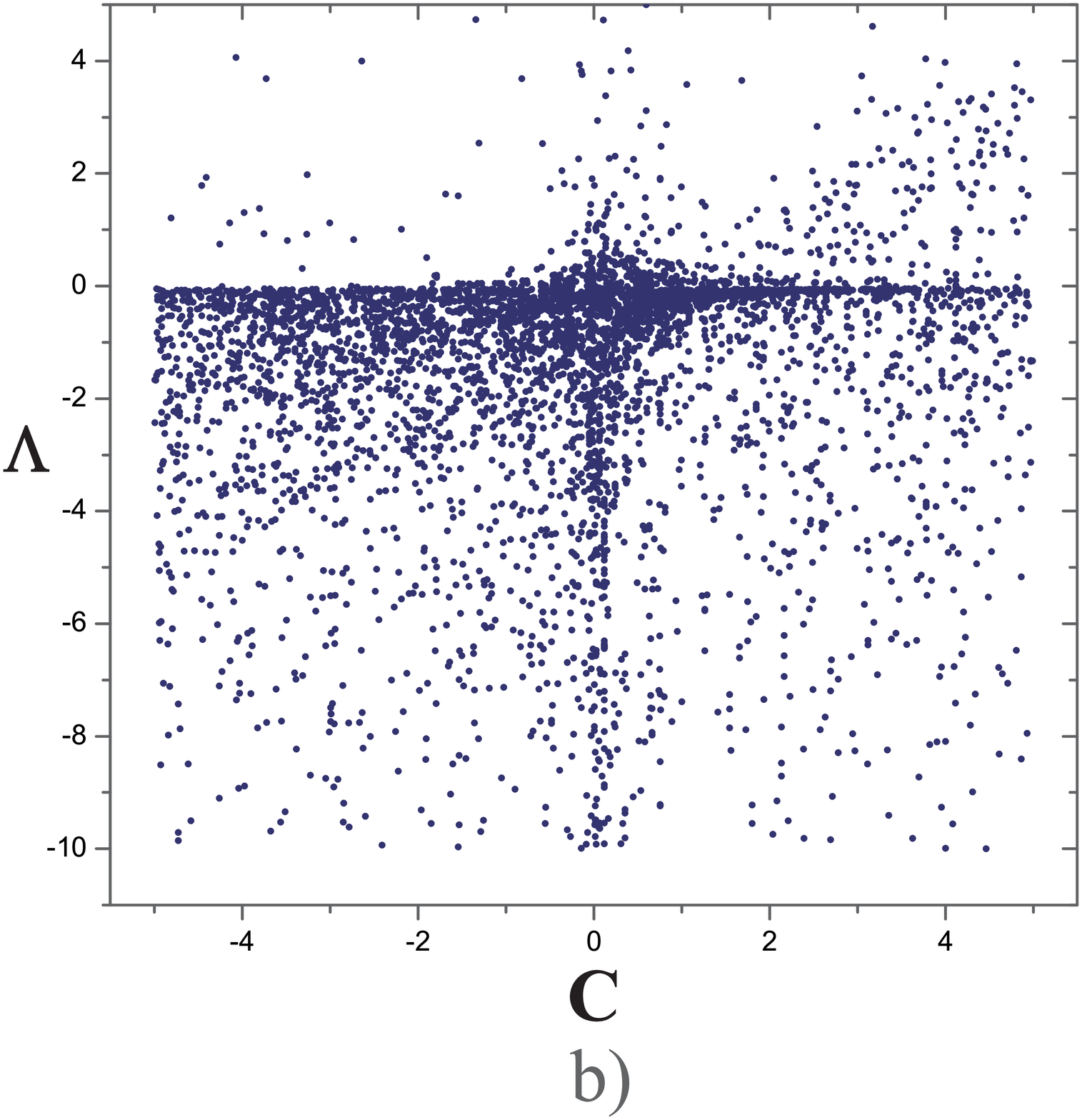}
\caption{Results of numerical computation: plane projections of the
points in the hyperspace $\big(C,C_1,C_2,\Lambda\big)$ corresponding
to solutions:\newline a) a projection on the plane $(C,C_1)$,\qquad
\qquad b) a projection on the plane $(C,\Lambda)$}
\label{points-planes}
\end{figure}

A set of initial parameters $\big\{C,C_1,C_2,\Lambda\big\}$ is
represented by a point in the parametric hyperspace
$\big(C,C_1,C_2,\Lambda \big)$. We shall refer to those parameter
sets that lead to observable constants (i.e. satisfy
\eqref{connect1}-\eqref{limit}) as \emph{solutions}. We are
interested in the density of solutions in the parametric
hyperspace. To assess the density visually we have to use its
plane projections -- for instance projections on planes
$\big(C_1,C_2\big)$ or $\big(C,\Lambda\big)$.
Fig.\,\ref{points-planes} presents the results of numerical
computation --
projections of a four-dimensional solution image on planes
$\big(C,C_1\big)$ and $\big (C,\Lambda\big).$

It could be seen that there are numerous solutions.  This means
that the observed low-energy physics is reproduced in a wide range
of initial parameters of our theory.

Recall that we took into account only the most simple
geometries of the hyperspaces undergoing reductions - only the
maximally symmetrical ones. Including other geometries into
consideration would yield additional solutions. If these solutions
form a continuum in the parametric space, any point in that space
(i.e. any set of initial parameters) would lead to the observable
physics.

\section{Discussion}

The prospective theory, called "the Theory of
Everything"\ (TOE), should solve the problem of the fine tuning of the universe. How are the initial parameters of TOE chosen and why do they allow very complex
structures to arise? A way to approach this problem was found by
proposing an existence of numerous vacua with different properties
-- a "landscape" in the modern terminology. The landscape, derived
in the scope of the string theory, is an important
advancement,
but the questions to be answered still remain. Is the concept of
'strings' indispensable or will the assumption of multiplicity of
dimensions alone suffice? How crucial are the values of the
initial parameters? What is the probability of getting to a
particular vacuum? Which additional values besides the metric do
we need to include?

Within the framework of the theory derived in this paper these
questions may be answered as follows:
\begin{itemize}
\item[-] For the low-energy 'landscape' to exist we need only to
assume the existence  of multiple dimensions. The rest of the
string theory tools as well as incorporation of additional
fields are not necessary. The cascade reduction mechanism
effectively produces the low-energy physics with various
parameters.

\item[-] A variation of the initial parameters does not affect
significantly the probability of a universe formation.

\item[-] Values of the Plank constant and the gravitational
constant differ in different universes and depend on the choice of
a reduction cascade for a given set of initial parameters. Those
constants are changing {with time} during the early stages
of the universe formation.
\end{itemize}

It has been shown that there are numerous values of initial
parameters of the theory which could be "connected by a cascade"\
with observed fundamental constants. Particular numerical values
of initial parameters are therefore not as important as it was
previously thought.

The 'landscape' concept implies that numerous low-energy
theories with various properties originate from a theory with
unique initial parameters. This paper is concerned with an
opposite scenario when the observed physics is derived from
numerous initial theories. This 'inverted landscape' model brings
into question the significance of a search for the unique
Lagrangian of TOE.

A set of all such values is rather large, although does not
constitute a continuum {in the parametric hyperspace}. This may be
attributed to the limited subset of all possible cascades that we
have studied -- only those composed of absolutely symmetric
compact spaces. Evidently an extension of this subset {to include
all possible geometries} will increase the number of acceptable
parameters.

\begin{center} \textbf{\textcolor[rgb]{0.2,.3,.3}{Conjecture}}  \end{center}
\vspace{-6mm}
\begin{quote}
\textit{For any given set of initial parameters there exists a cascade of reductions leading from a multidimensional space to a universe of our type.}
\end{quote}

The credibility of this hypothesis will be fully examined in the future research.

\vspace{5mm}

\subsection*{Acknowledgements.} We are grateful to K.A. Bronnikov for helpful
discussions. This work was supported in part by the Russian
Foundation for Basic Research (project no. 09-02-00677-a)

\end{document}